\documentclass[12pt]{article}
\usepackage{epsfig}
\usepackage{amssymb,graphicx}
\def\be{\begin{equation}}
\def\ee{\end{equation}}
\def\ba{\begin{array}{c}}
\def\ea{\end{array}}

\newcommand{\bea}{\begin{eqnarray}}
\newcommand{\eea}{\end{eqnarray}}

\begin{document}

\begin{center}

{\Large \bf {

 Relocalization switch in a triple quantum dot molecule in 2D

 }}

\vspace{13mm}

\vspace{3mm}

\begin{center}

\textbf{Miloslav Znojil}


\vspace{1mm}

The Czech Academy of Sciences, Nuclear Physics Institute,
 Hlavn\'{\i} 130,
250 68 \v{R}e\v{z}, Czech Republic

\vspace{0.2cm}

 and

\vspace{0.2cm}

Department of Physics, Faculty of Science, University of Hradec
Kr\'{a}lov\'{e},

Rokitansk\'{e}ho 62, 50003 Hradec Kr\'{a}lov\'{e},
 Czech Republic

\vspace{0.2cm}

{e-mail: znojil@ujf.cas.cz}

%
%
%
%

\end{center}

\vspace{3mm}

\end{center}

\subsection*{Keywords:}

.

 coupled quantum dots in 2D;

 probability density localization;

 non-separable polynomial model;

 quantum catastrophe control;

\subsection*{PACS number:}

PACS 03.65.Ge - Solutions of wave equations: bound states

\subsection*{Abstract}

A symmetric chain of three quantum dots
(i.e., one of the simplest quantum
dot molecules)
is constructed using
a three-parametric non-separable version of an
asymptotically separable sextic polynomial potential $V(x,y)
= (x^2+y^2)^3+\ldots$.
The probability density $|\psi(x,y)|$
(admitting either the central or off-central dominance)
is assumed measured. A dynamical regime is found with an
enhanced sensitivity
of the central -- off-central transition
to the parameters.
Quantitatively, the possibility of
control of such a switch {\it alias\,}
``relocalization catastrophe'' is
illustrated non-numerically.

\newpage

\section{Introduction}

In the unitary version of quantum mechanics of confined systems one encounters
a wide methodical gap between their one-dimensional subfamily
described by Schr\"{o}dinger equation
 \be
 \left[- \frac{\hbar^2}{2\mu}\,\frac{d^2}{dx^2} +
 V(x)
 \right ] \psi_n (x) =E_n\,
 \psi_n (x)
  \,, \ \ \ \ \
 n=0, 1, \ldots\,
 \label{1dse}
 \ee
and various more general scenarios. They
may be sampled by the very next 2D family described by
partial differential
Schr\"{o}dinger equation
 \be
 \left[- \frac{\hbar^2}{2\mu_x}\,\frac{d^2}{dx^2}
 - \frac{\hbar^2}{2\mu_y}\,\frac{d^2}{dy^2}+
 V(x,y)
 \right ] \psi_m (x,y) =E_m\,
 \psi_m (x,y)
  \,, \ \ \ \ \
 m=0, 1, \ldots\,.
 \label{2dse}
 \ee
The gap involves not only constructive mathematics (which is
perceivably more complicated in the latter case) but also
experimental physics where, on the contrary, the realization of the
confinement is usually more challenging in the one-dimensional setup.

We believe that the
transition from the most popular 1D Eq.~(\ref{1dse}) to the
non-separable two-dimensional bound-state problem (\ref{2dse}) may
be more straightforward than expected.
In an analysis of this possibility we feel
motivated by the results of our recent paper \cite{arpot}.
Indeed, for multiple polynomial multi-well
one-dimensional potentials $V(x)$ we managed to localize there
the existence of experimentally interesting
bifurcations of probability densities $|\psi_n (x)|$.
Moreover, we revealed that the theoretical prediction of such a
``quantum catastrophic'' phenomenon
may remain non-numerical or, at worst,
semi-numerical.

In our present letter we intend to show that
a similar successful determination of the
sets of the critical coupling constants
may be also achieved and performed in 2D.

\section{Classical versus quantum catastrophes}

Stable stationary equilibria of a classical dynamical system ${\cal
S}$ may be identified with the minima of an {\it ad hoc\,} potential
$V=V_{\cal S}(x)$ or $V=V_{\cal S}(x,y)$, etc. The dependence of
properties of these equilibria on parameters was given an elementary
qualitative classification by R. Thom \cite{webcat}. In applications
\cite{Thom} he proposed to speak about the ``fold catastrophe''
(using the benchmark choice of $V^{(fold)}(x)=x^3+ax$ characterized
by a bifurcation of equilibria at $a=0$), about the ``cusp
catastrophe'' (using $V^{(cusp)}(x)=x^4+ax^2+bx$),
etc. Naturally, the transparent
geometric nature as well as the conceptual simplicity made this
classification scheme useful in diverse areas ranging from the
evolution scenarios in biology \cite{revcat} up to various
phenomenological aspects of quantum chemistry \cite{Thomchem}.

In \cite{arpot} we pointed out that after a tentative transition to
quantum systems ${\cal Q}$ and to the related bound-state
Schr\"{o}dinger Eqs.~(\ref{1dse}) or~(\ref{2dse}), the spectrum need
not remain discrete (and/or bounded below) so that the paradigm
(i.e., the Thom's classification scheme) must be amended. We
emphasized that after quantization one has to omit, first of all,
all of the ``non-confining'' Thom's potentials, i.e., not only the
above-mentioned odd-parity function $V^{(fold)}(x)$ but also another
one-dimensional potential $V^{(swallowtail)}(x)=x^5+ax^3+bx^2+cx$ as
well as all of the three two-dimensional ``umbilic'' benchmarks
$V(x,y)$.

As a result, the Thom's menu of seven basic dynamical scenarios of
an immediate classical applicability degenerates to the mere doublet
of one-dimensional quantum potentials in Eq.~(\ref{1dse}) formed by
the above-mentioned two-parametric function $V^{(cusp)}(x)$ and/or
by the four-parametric function
 \be
 V^{(butter\!fly)}(x)=x^6+ax^4+bx^3+cx^2+dx\,.
 \label{butt}
 \ee
Moreover, in the context of quantum physics one also has to consider
the effect of tunneling. At a negative $a<0$ in $V^{(cusp)}(x)$, for
example, the particle may be detected localized near {\em both\,} of
the minima separated by the barrier. One can conclude that only the
choice of $V^{(butter\!fly)}(x)$ makes good sense and can be made
responsible for simulation of a nontrivial observable quantum
catastrophe.

In \cite{arpot} we analyzed the butterfly model (\ref{butt}) and
emphasized the existence of a change-of-localization catastrophe
during which the dominance of the central local minimum at $x=0$
becomes replaced by the topologically non-equivalent dominance of
the two ``remote'' minima. Under a simplifying assumption of
vanishing $b=d=0$ we were able to interpret such a ``quantum
catastrophe'' as a phenomenon of spatial bifurcation of the
probability density $|\psi(x)|$.

In applications, unfortunately, the existence of the formal theoretical and
mathematical merits of the one-dimensional model 
may be found accompanied 
by the emergence of a certain discomfort in experimental setting.
Indeed, the schematic one-dimensional, $y-$independent
interactions $V(x,y)=V^{(butter\!fly)}(x)$ 
(admitting the free motion
along the $y-$axis)
have to be realized,
in the laboratory, typically, by the parallel waveguides which have 
to be very long (i.e., in principle ``infinitely'' long).  

Naturally, one of the most natural ways of suppression of such 
an obvious phenomenological disadvantage
can and should be sought via a transition to the genuine two-dimensional models. 
In such a direction of research we were mainly encouraged 
by the observation that after an extension of our study \cite{arpot} to 2D,
the underlying Schr\"{o}dinger Eq.~(\ref{2dse})
happened to be, in the dynamical regimes of interest, by far not that
difficult to solve and analyze, in spite of its non-separability. 

In our
present letter a certain user-friendly family of models based on 2D potentials
$V(x,y)$ will be proposed and considered, therefore.
Briefly, these models may be characterized
by the 
survival of the two basic
mathematical merits of their 1D predecessors.
The first one is mathematical,
lying in the elementary non-numerical feasibility of an
exact localization of the 
pronounced coupling-dependent
minima and maxima of the potentials in both of their 
1D and 2D exemplifications.
The second key merit of the present 2D extension of the family
of tractable models is more pragmatic. 
Its essence (to be explained below) lies in the purely
empirical observation that 
the above-mentioned determination of the
``quantum
catastrophic'' dynamical regime becomes facilitated
by the unexpectedly user-friendly nature of the 
bifurcation
phenomenon which appears ``abrupt'', 
restricting 
the breakdown of the non-numerical tractability
of the 
transition process to a  very narrow interval of the couplings 
in {\em both\,} of the 1D and 2D models.

\section{Quantum bound states in 2D}

\subsection{Asymptotically separable toy-model potential and its extremes}

Although the Thom's classical theory does not offer, in two
dimensions, any obvious methodical guidance, we intend to use a
modified idea of Ref.~\cite{arpot}. By itself, it was inspired by
the Arnold's extension \cite{Arnold} of the Thom's list of the
specific one-dimensional benchmarks $V(x)$ to the higher-degree
polynomials. This means that in what follows we shall also admit an
arbitrary polynomial form of $V(x,y)$. Still, for the sake of
brevity we shall only restrict our attention to the sufficiently
elementary three-parametric two-dimensional model of the strictly
confining polynomial form
 \be
  V(x,y)=
  r^6
  -A^2r^4+(B^2+C^2y^2)r^2\,,
  \ \ \ \ \
  r^2=x^2+y^2\,.
  \label{huja}
 \ee
At $y=0$ this function formally degenerates to the first nontrivial
two-barrier model of Ref.~\cite{arpot} so that we may reparametrize
its first two coupling constants,
 \be
 A^2=3(\alpha^2+\beta^2)\,,
 \ \ \ \
 B^2=3\alpha^2\gamma^2\,,
 \ \ \ \
 \gamma^2=\alpha^2+2\,\beta^2
 \,.
 \ee
According to section \# 3 in \cite{arpot} the three local minima of
function (\ref{huja}) do exist and do lie at $r=0$ (with vanishing
value of $V(0,0)=0$) and at $(x,y)=(\pm \gamma,0)$ (where we have
$V(\pm \gamma,0)=(\alpha^2-\beta^2)\gamma^4$). The height of the
local saddle-point-shaped barriers is given by formula $V(\pm
\alpha,0)=\alpha^4(\beta^2+\gamma^2)$.

For the large values of parameters $\alpha$ and $\beta$ (with even
larger $\gamma=\gamma(\alpha,\beta)$) the barrier is high so that
the three local minima  (all occurring at $y=0$) are well separated.
In such a dynamical regime our present 2D quantum model may be perceived 
as qualitatively fully analogous to its 1D predecessors of Ref.~\cite{arpot}.
Not quite expectedly, also the leading-order
determination of the low-lying spectra of quantum bound-state energies
will be shown to remain fully analogous and manifestly non-numerical.

\subsection{The mechanism of relocalization of 2D probability densities}

The ground state wave function may be found localized either
near $x=r=0$ (provided that we have $\alpha \gg \beta$), or near the two
separate centers with $x=\pm \gamma$ (provided that $\alpha \ll
\beta$). A sudden ``relocalization'' {\it alias\,} catastrophic
transition between these two topologically non-equivalent
distributions of probability of finding the particle should finally
be observed at $\alpha \approx \beta \gg 1$.

\subsubsection{Dynamical regime of centrally localized probability densities}

\begin{figure}[h]                    
\begin{center}                         
\epsfig{file=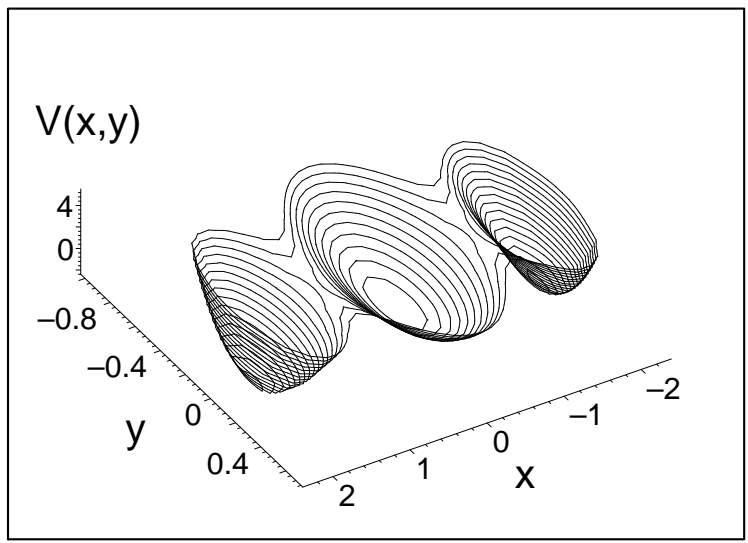,angle=270,width=0.56\textwidth}
\end{center}    
\vspace{2mm} \caption{Sample potential $V(x,y)= \left(
{x}^{2}+{y}^{2} \right) ^{3}- 6.6\, \left( {x}^{2}+{y}^{2} \right)
^{2}+ \left( 10.2+9\,{y}^{2} \right) \left( {x}^{2}+{y}^{2}
 \right)$ near its three local minima. Parameters are $\alpha^2=1$,
$\beta^2=1.2$ and $\gamma^2=3.4$, the two outer minima are $V(\pm
\gamma,0)=-2.312$, while the heights of the two saddle-point
barriers are $V(\pm \alpha,0)=3.4$.
 \label{fione}
 }
\end{figure}

We have to remind 
the readers here that as long as we speak about the qualitative 
features of a non-trivial, 2D quantum system, the determination of the probability
densities $|\psi(x,y)|$ obtainable from the manifestly 
non-separable Sch\"{o}dinger Eq.~(\ref{2dse})
is a truly challenging task requiring an application of the 
standard numerical construction techniques in general. Fortunately,
once we restrict attention to the domain of large parameters
$\alpha$ and $\beta$, our qualitative understanding of the 
dynamics becomes thoroughly simplified, based on the generic
graphical illustration of the shape of $V(x,y)$ as provided by
Fig.~\ref{fione}.


In the picture we see that in the vicinity of the origin the shape 
of the potential
acquires
the elementary analytic representation given directly by
Eq.~(\ref{huja}),
 \be
 V(x,y)= 3\,\alpha^2\gamma^2\,r^2
 + {\cal O}(r^4)\,.
 \ee
Whenever we have, by our overall assumptions, 
$\alpha \gg \beta \gg 1$, the value of potential $V(0,0)=0$
will represent a pronounced  global minimum of its shape. 

The most important feature of the shape of  $V(x,y)$
near this minimum is its central symmetry.
The potential becomes very well approximated
by the exactly solvable harmonic-oscillator well.
Thus, one immediately obtains
the leading-order formula for energies
 \be
 E_n^{(0)}=2\,\sqrt{3}(2n+1)\alpha\gamma + \ldots\,
 \ \ \ \ n = 0, 1, \ldots\,.
 \label{centr}
 \ee
This formula may be then used as an approximation of the low-lying part
of the spectrum of the bound-state energies when calculated in the units
such that ${\hbar^2}/({2\mu_x})={\hbar^2}/({2\mu_y})=1$.

\subsubsection{Dynamical regime of the centrally suppressed probability densities}

Due to the polynomial form of the potential we may also easily
derive the harmonic-oscillator approximation of the potential near
its other two local minima lying at $x=\pm \gamma$,
 \be
 V(x,y)=(\alpha^2-\beta^2)\gamma^4 + 12\,\beta^2\gamma^2 \xi^2+
 C^2\gamma^2y^2 + \ldots\,,
 \ \ \ \ \
 x=\pm \gamma+\xi\,.
 \ee
Again, under the alternative assumption that these minima are global
and pronounced (i.e., that $1 \ll \alpha \ll \beta$), we arrive at
the alternative leading-order formula for the corresponding
twice-degenerate (i.e., parity-degenerate) low-lying spectrum,
 \be
 E_{m,k}^{(\pm)}=
 (\alpha^2-\beta^2)\gamma^4+
 2\,\sqrt{3}(2m+1)\beta\gamma+ C\,\gamma\,(2k+1)
 + \ldots\,,
 \ \ \ \ m,k = 0, 1, \ldots\,.
 \label{offcentr}
  \ee
Naturally, the dominance of these energies in the low-lying spectrum
of the whole system is given by the comparative smallness of $\alpha$
which still, by our assumptions, remains also very large in its absolute 
value. 

Now we are prepared to compare the two 
alternative approximate formulate for the quantum
ground state
energies. Firstly,
in the dominant$-\alpha$
regime we see that the
ground-state energy (\ref{centr}) is $\beta-$independent, positive and
only ``slowly'' growing as a quadratic 
function of $\alpha$, $E_n^{(0)}\approx E_n^{(0)}(\alpha) = {\cal O}(\alpha^2)$.
In contrast,
in the dominant$-\beta$
regime 
the shape of the $\alpha-$ and $\beta-$dependent dominant-energy
surface $E_{m,k}^{(\pm)}=E_{m,k}^{(\pm)}(\alpha,\beta)$
of Eq.~(\ref{offcentr}) is, as a function of
$\beta$ studied at a fixed $\alpha$, deeply negative, very steep 
($E_{m,k}^{(\pm)}(\alpha,\beta)={\cal O}(\beta^6)$)
and, in the vicinity of the end-of-the-domain limit $\beta\to \alpha^+$, 
even sign-changing (mainly due to the second, correction term ${\cal O}(\beta^2)$ 
which is positive). 

Both of these parameter-dependence tendencies are monotonous.
This is their important property which
implies that the system in question will
necessarily pass through its
relocalization critical point at which the separate approximate
subspectra (\ref{centr}) and (\ref{offcentr}) overlap. 
Due to the extremely 
steep, ${\cal O}(\beta^6)$ shape of the latter, $\beta-$dominant energy curve,
the ``critical'' interval of  $\beta \approx \alpha \gg 1$ 
(in which 
one would have to use brute-force numerical methods since
both of our approximations (\ref{centr}) and (\ref{offcentr}) 
cease to be valid) remains extremely narrow.
The change of the localization of the densities $|\psi(x,y)|$
becomes abrupt.

\subsubsection{Quantum relocalization phase transition}

In the light of our preceding considerations, the 
instant of the change of the 
type of dominance of densities $|\psi(x,y)|$ may
be determined very easily.
On our present level of precision it is sufficient to
determine the instant of the intersection
and coincidence of the two alternative candidates
for the ground-state energy,
 \be
 E_0^{(0)}(\alpha)\approx E_{0,0}^{(+)}(\alpha,\beta)\,.
 \label{rovn}
 \ee
Up to the higher-order corrections the latter relation
determines the relocalization catastrophe (RC).

After a detailed inspection of our formulae we reveal that 
among the three coupling constants in the potential
the two (viz., $A$ and $B$) were assumed, from the very beginning, 
large.
For this reason,
a decisive role 
determining the RC transition remains
to be played by the third coupling
$C$.
In fact, this parameter
measures the asymmetry of the potential 
so that its critical value $C=C^{(RC)}$
can also be treated
as a
function of the two other, freely variable dynamics-determining
quantities $\alpha\gg 1$ and $\beta\gg 1$.

After an elementary algebra we obtain, from (\ref{rovn}), 
our ultimate explicit formula
 \be
 C^{(RC)}=\left [
 (\alpha+\beta)\gamma^3-2\sqrt{3}
 \right ](\beta-\alpha)
  \,.
  \label{inn}
 \ee
Its last-step simplification
 $$
 C^{(RC)}=
 (\beta^2-\alpha^2)\gamma^3 + {\rm corrections}
 $$
characterizes the 
dominant
RC relationship resulting from the observation that in
(\ref{inn}) the first term in the square brackets is dominant. Thus,
we must have $\beta>\alpha \gg 1$, i.e., say,
$\beta^2=\alpha^2+q\alpha$ with a bounded auxiliary factor $q =
{\cal O}(1)$. Nevertheless, an additional, very reasonable
requirement of a parallel  boundedness of $C^{(RC)}=\nu \alpha$
(with, say, $\nu = {\cal O}(1)$) leads to the amended estimate
$q=\nu/(\sqrt{27}\alpha^3)$ so that in the critical regime the
admissible difference between $\beta$ and $\alpha$ remains extremely
small,
 $$
 \beta^2-\alpha^2=q\alpha={\cal O}(\alpha^{-2})\,.
 $$
This observation only reconfirms that, as expected, 
the process of the RC catastrophe is truly
``abrupt''.

%
%

%
%
%
%
%
%
%
%
%
%
%
%
%
%
%
%
%
%
%

%
%
%
%

\section{Outlook}

In the quasi-classical limit of vanishing $\hbar \to 0$ (or, if you
wish, of infinite mass $\mu \to \infty$), all of the low-lying
quantum bound states described by Schr\"{o}dinger Eqs.~(\ref{1dse})
or (\ref{2dse}) become degenerate and localized strictly at the
dominant minima of the potential. This may be perceived as a certain
unidirectional
correspondence between the genuine quantum equilibria
and the Thom's classical
(i.e.,
non-quantum) concepts of bifurcations
{\it alias\,} catastrophes.

Formally speaking, the correspondence is not one-to-one because
the classical theory admits also the potentials
(called, in this context, Lyapunov functions) which are {\em not\,}
asymptotically confining. In quantum world, on the contrary, one can easily
construct various non-analytic potentials and systems exhibiting a
catastrophic bifurcation behavior. Indeed, quantum Hamiltonians need not
necessarily be defined in the Thom-inspired differential-operator
form at all.

Our present letter demonstrated our belief that the parallels between
the classical and quantum theory
of catastrophes may be strengthened, anyhow. We
followed the idea of searching for them
under the
reasonable
assumption of
using the most elementary
polynomial
forms of the local
potential functions.
In such an overall research
project
there emerged several
key mathematical challenges.
The main one
may be seen in the fact that
immediately after one moves from the one-dimensional models of
paper \cite{arpot} to the much broader family of
$D-$dimensional quantum systems with $D \geq 2$, the generic loss of
the separability of the underlying Schr\"{o}dinger equation seems to
imply the necessity of a purely numerical localization and
description of the catastrophes.
In this sense our present results might
be perceived as encouraging.

Let us also add that during our project we felt guided by the
not quite expected mathematical friendliness of the
one-dimensional constructions in Ref.~\cite{arpot}. Indeed, in that paper we
managed to circumvent the two mutually interrelated difficulties of
our catastrophe-simulation model-building, viz.,
the smeared, non-abrupt nature of the
bifurcations
(which is, in a way, typical
for quantized systems),
and an apparently purely numerical character of the
solutions of Schr\"{o}dinger equations,
especially in the higher-degree
polynomial-potential models.

In our present letter we came to a few equally encouraging observations.
Our success has been
achieved via our restriction of attention just to the
very next, two-dimensional
quantum systems, and to
the
dynamical
regime characterized by the ``sufficiently pronounced'' local minima
of $V(x,y)$ separated by the ``sufficiently high'' barriers.
In this sense a similar success to be achieved,
in the nearest future,
at $D=3$ etc., need not be excluded.

For the sake of brevity we picked up just one of the most elementary
nontrivial models, viz.,
the two-dimensional butterfly interaction (\ref{huja}).
We were fortunate in revealing that even this next to elementary
potential already
exhibited many of the
desirable phenomenological properties.
Besides that, several
challenging mathematical
features of the model
(like its non-separability)
were found tractable.
This proved promising and welcome, in particular, in the light of their
quantum-theoretical relevance
sampled, first of all, by the
existence of the probability density relocalization phenomenon.

Marginally let us finally re-emphasize that in the phenomenologically most
interesting regime of the competition for dominance between the
multiple (i.e., in our model, three) pronounced local minima
(approximated, with good precision, by the solvable
harmonic-oscillator wells, and connected via the quantum tunneling
of course) the determination of the instant of the ``catastrophe''
(i.e., of an abrupt change of the topology of the probability
density $|\psi(x,y)|$) remained purely non-numerical.

\section*{Acknowledgments}

The author acknowledges the support by the
 Faculty of Science of the University of Hradec
Kr\'{a}lov\'{e} and, in particular, by the
Excellence project 2212 P\v{r}F UHK 2020.


\newpage

\begin{thebibliography}{00}



\bibitem{arpot}
M. Znojil, Ann. Phys. 413, 168050 (2020).


\bibitem{webcat}
https://en.wikipedia.org/wiki/Catastrophe\_theory


\bibitem{Thom}
R. Thom, Structural Stability and Morphogenesis: An Outline of a
General Theory of Models. Addison-Wesley, Reading,  1989.


%
%
%
%
%
%
%
%
%
%

%
%
%
%


\bibitem{revcat}
J. Poston and I. Stewart, Catastrophe Theory and Its Applications,
Pitnam, London, 1978;

\bibitem{Thomchem}
X. Krokidis, S. Noury and B. Silvi, J. Phys. Chem 101, 7277 (1997).
%
%
%
%
%
%
%
%
%
%
%
%

%


\bibitem{Arnold}
V. I. Arnold,
Catastrophe Theory. Springer-Verlag,  Berlin, 1992.

%


\end{thebibliography}
\end{document}